\documentclass[aps,prl,twocolumn,superscriptaddress,amsmath,amsfonts,amssymb,preprintnumbers]{revtex4-1}
\usepackage[T1]{fontenc}
\usepackage{wasysym}
\usepackage[latin1]{inputenc}
\usepackage{graphicx}
\usepackage[colorlinks,plainpages=false,linkcolor=black,urlcolor=blue,citecolor=black,pdfpagemode=UseNone,pdfstartview=FitBH]{hyperref}

\makeatletter
\renewcommand{\@evenfoot}{\hfill\raisebox{-3em}{\bf\thepage}\hfill}
\renewcommand{\@oddfoot}{\hfill\raisebox{-3em}{\bf\thepage}\hfill}
\makeatother

\begin{document}

\title{Extreme thermopower anisotropy and interchain transport in the quasi-one-dimensional metal Li$_{0.9}$Mo$_6$O$_{17}$}

\author{J.~L.~Cohn}
\email[Corresponding Author: ]{cohn@physics.miami.edu}
\affiliation{Department of Physics, University of Miami, Coral Gables, FL 33124}

\author{S.~Moshfeghyeganeh}
\affiliation{Department of Physics, University of Miami, Coral Gables, FL 33124}

\author{C.~A.~M. dos Santos}
\affiliation{Escola de Engenharia de Lorena - USP, P. O. Box 116, Lorena-SP, 12602-810, Brazil}

\author{J.~J.~Neumeier}
\affiliation{Department of Physics, Montana State University, Bozeman, Montana 59717}

\begin{abstract}
Thermopower and electrical resistivity measurements transverse to the conducting chains of
the quasi-one-dimensional metal Li$_{0.9}$Mo$_6$O$_{17}$ are reported in the temperature range
5~K $\leq T\leq$ 500~K. For $T\geq 400$~K the interchain transport is determined by thermal
excitation of charge carriers from a valence band $\sim 0.14$~eV below the Fermi level, giving rise
to a large, $p$-type thermopower that coincides with a small, $n$-type thermopower along the chains.
This dichotomy -- semiconductor-like in one direction and metallic in a mutually perpendicular direction --
gives rise to substantial transverse thermoelectric (TE) effects and a transverse TE figure of merit
among the largest known for a single compound.
\end{abstract}

\maketitle
\clearpage

Conducting materials with highly anisotropic Seebeck coefficients (thermoelectric powers or TEPs) are potentially
useful in transverse thermoelectric applications for energy detection and cooling \cite{Goldsmid,Lengfellner,GraysonPRL}.
Bulk conductors for which the TEPs in different crystallographic directions have opposite signs and yield
a large magnitude for their difference ($\Delta S\geq 200\ \mu$V/K) are quite rare \cite{Ong,GraysonPRL}, thus recent developments
have focused on artificial synthesis of stacked bulk materials \cite{Goldsmid,Lengfellner} or semiconductor
heterostructures \cite{GraysonPRL} to achieve large Seebeck anisotropy.
Here we present transport measurements on the quasi-one-dimensional (q1D) metal, Li$_{0.9}$Mo$_6$O$_{17}$
known as ``lithium purple bronze'' (LiPB), that reveal a surprisingly simple mechanism for extreme Seebeck
anisotropy in a bulk conductor. Direct electron transfer between the q1D metallic chains of this material
is sufficiently weak that interchain transport above 400 K is predominated by thermal activation of
valence band states ($\sim 0.14$~eV below $E_F$), yielding a large, {\emph p}-type interchain Seebeck
coefficient that coexists with {\emph n}-type metallic behavior
confined along the q1D chains. A substantial transverse Peltier effect is demonstrated.
These ingredients may exist in other materials or might possibly be engineered to develop
transverse thermoelectrics based on a single compound.

A resistivity that is metallic at low-temperature
and decreases anomalously at high temperatures is a ubiquitous characteristic of
transport transverse to the planes or chains of many q2D \cite{graphite,SrRuO4,NaCoO4,k-(BEDT-TTF)2Cu(SCN)2} and q1D
metals \cite{Pr124,BaRu6O12,KRu4O8}, respectively.  It is generally accepted that this behavior is due to the onset of
an additional conduction mechanism in parallel with band transport, possibly related to interplane or
-chain defects (e.g. resonant tunneling) \cite{GutmanMaslov}.

Much less is known about the TEP transverse to the planes or chains of such materials, partly because TEP measurements are
difficult to perform in small single crystals for which the transverse transport directions have very small dimensions (e.g., thin
platelet or needle-like habits). In the few compounds where transverse TEP measurements have been
reported \cite{Sb2Te3+Bi2Te3,Graphites,TMTSF2X,Y123,Bi2212,Ca3Co4O9}, high anisotropy has not been observed.

Li$_{0.9}$Mo$_6$O$_{17}$ known as ``lithium purple bronze'' (LiPB), is a low-temperature superconductor ($T_c\approx 2$~K)
first synthesized and studied in the 1980s \cite{PBReview,oldwork1,BandStructure1}. It has attracted interest more recently for its quasi-one
dimensionality and Luttinger-liquid candidacy \cite{RecentAllen,BandStructure2,Optical,NeumeierPRL,RecentTheory1,RecentTheory2,NussAichhorn}.
Crystal growth \cite{oldwork1,NeumeierPRL} and transport properties along the chains (crystallographic {\emph b} axis) for
crystals similar to those discussed here have been presented elsewhere \cite{XRD,Nernst}. The resistivity anisotropy of LiPB is
approximately \cite{Mercure} $\rho_b:\rho_c:\rho_a= 1:80:1600$. Single-crystal specimens were oriented by x-ray diffraction and
cut/polished into thin rectangular plates with the thinnest dimension (along the \emph{a} axis) typically 40-80 $\mu$m.
The \emph{bc}-plane dimensions were typically $0.4\times 1.0$~mm with the longest dimension coinciding
with the transport axis (\emph{b} or \emph{c} direction). Electrical contacts were made with Au leads attached with silver
epoxy. Current contacts covered the specimen ends and voltage contacts encircled the crystals across both large faces and the
sides. For thermopower measurements, specimens were suspended from a Cu heat sink with silver epoxy and affixed with a heater and
25-$\mu$m-diameter differential chromel-constantan thermocouple, both attached with stycast epoxy. Separate radiation-shielded
vacuum probes were employed for the cryogenic and high-$T$ (> 320~K) measurements.

\begin{figure}[t]
\includegraphics[width=3.125in,clip]{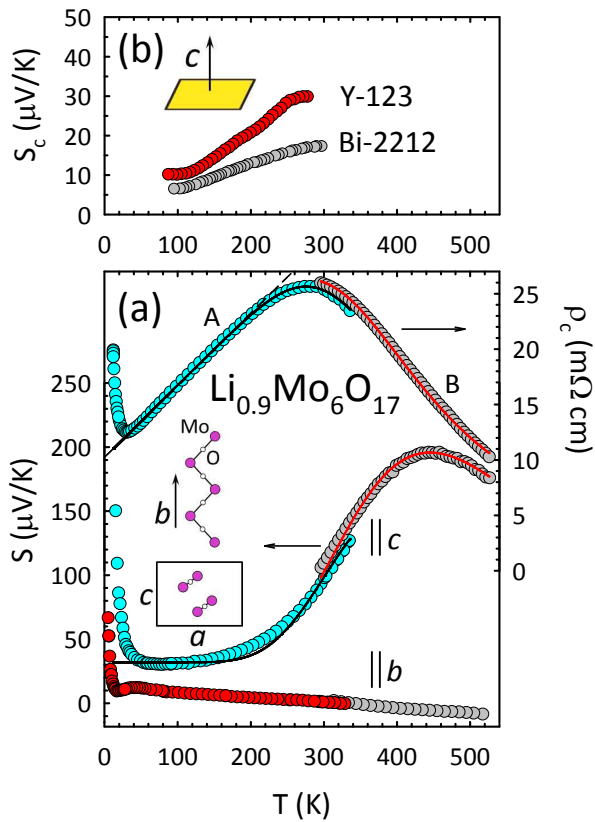}
\caption{(color online) (a) LiPB interchain ({\emph c}-axis)
thermopowers (left ordinate) and resistivities (right ordinate)
for two crystals (labeled A and B), and intrachain ({\emph
b}-axis) thermopowers for two different crystals. Solid curves
through the \emph{c}-axis data are fits to the parallel conduction
model discussed in the text; parameters for the semiconducting
component are listed in Table~I. Dashed line is a
linear-least-squares fit to the low-$T$ $\rho_c(T)$ (see text).
Inset: orientation of the crystallographic axes with respect to
the q1D Mo-O chains (2 per unit cell). (b) interplane ({\emph
c}-axis) thermopowers for YBa$_2$Cu$_3$O$_7$ (Y-123,
Ref.~\onlinecite{Y123}) and Bi$_2$Sr$_2$CaCu$_2$O$_8$ (Bi-2212,
Ref.~\onlinecite{Bi2212}).} \label{Fig1}
\end{figure}

Figure~\ref{Fig1}~(a) shows for two crystals the interchain resistivity and TEP, $\rho_c(T)$ and $S_c(T)$, along with the
intrachain TEP, $S_b(T)$, for two different crystals \cite{XRD}. Additional {\emph c}-axis data for two more crystals
can be found in the Supplementary Material \cite{SM}. The increase in all three coefficients below 30~K has been
discussed extensively elsewhere \cite{XRD} and may be associated with localization,
dimensional crossover or the development of unconventional (e.g., electronically-driven) charge density-wave order \cite{Optical,NeumeierPRL}.
The focus of the present work is the interchain TEP in the region $T > 40$~K where it rises sharply with increasing $T$,
coincident with a deviation of $\rho_c(T)$ from it's low-$T$, linear-$T$ behavior (dashed line, Fig.~\ref{Fig1}).  The intrachain TEP is
linear-in-$T$, modest in magnitude, and becomes negative above 300 K, consistent with electron-like carrier diffusion as noted previously \cite{XRD},
and extended here to 520~K.

Several features of $S_c(T)$ are noteworthy. It remains positive throughout the temperature range. In the linear-$T$
regime of $\rho_c$ (40 K$< T<$ 140 K), $S_c$ is nearly $T$-independent at  $\sim 32\ \mu V/K$ and essentially the
same for all crystals measured. Near the maximum in $S_c$ at $T\simeq 440$~K, $\Delta S=S_c-S_b\geq 200\ \mu$V/K.
The interchain transport in LiPB is incoherent, the metallic character of $\rho_c$ in the lower-$T$ regime
likely reflecting the intrachain scattering rate \cite{KumarJayan}, consistent with very weak and
indirect interchain hopping \cite{NussAichhorn}.
The nearly constant TEP at low-$T$ is a characteristic of narrow-band hopping \cite{Fisher}.
As for the increase in $S_c$ to very large values at higher $T$, it is instructive to compare with the behavior found for
the interplane ({\emph c}-axis) TEPs of the q2D metals \cite{Y123,Bi2212}, YBa$_2$Cu$_3$O$_7$ and Bi$_2$Sr$_2$CaCu$_2$O$_8$ [Fig.~\ref{Fig1}~(b)].
The TEPs of the latter materials, truncated by the onset of superconductivity at low-$T$, also rise for
$T\gtrsim 120$~K, but their overall increases ($\lesssim 20\ \mu$V/K)
are substantially smaller than for LiPB and a tendency toward saturation is evident at the highest $T$.
Their upturns are plausibly attributed to the onset of interplane tunneling, though theoretical
work \cite{Schofield} has not yet treated the TEP within a model that
incorporates resonant tunneling through defects \cite{GutmanMaslov}. The upturn in the interchain TEP for LiPB is
qualitatively and quantitatively different.
\begin{figure}[t]
\includegraphics[width=2.95in,clip]{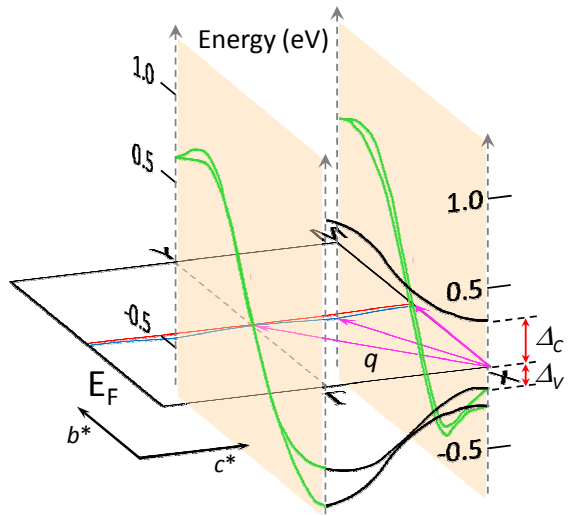}
\caption{(color online) Bands along main symmetry directions
(green: along {\emph b}$^*$, black: along {\emph c}$^*$) and
projected Fermi surface (red and blue curves) in the {\emph
b}$^*${\emph c}$^*$ plane for LiPB (adapted from
Ref.~\onlinecite{NussAichhorn}). $\Delta_V$ and $\Delta_C$ are
excitation energies for valence- and conduction-band states,
respectively, dispersing along {\emph c}$^*$ near the X-point of
the Brillouin zone. Indirect transitions between these states and
the Fermi surface are possible through thermal excitation and
absorbtion of a phonon with sufficient momentum along {\emph
b}$^*$ (wavevectors for excitation of valence-band states are
represented by pink arrows).} \label{Fig2}
\end{figure}

We propose that this difference has its
origin in the LiPB band structure which is distinguished from these other compounds by the presence of
valence and conduction bands in close energy proximity to $E_F$ and with sufficient dispersion for interchain
momentum so as to become increasingly important for interchain transport with increasing $T$.
Figure~\ref{Fig2} shows the calculated band structure \cite{BandStructure1,BandStructure2,RecentTheory2,NussAichhorn}
within the {\emph b}$^*$-{\emph c}$^*$ plane (dispersion along {\emph a}$^*$ is negligible). The projected
Fermi surface (blue and red curves) is also shown.
Electrons in valence bands dispersing along $c^*$ (black curves) can make indirect,
interband transitions to states at $E_F$ through thermal excitation ($\Delta_V$, Fig.~2) and
absorption of phonons with sufficient momenta ($q$) along the {\emph b}$^*$ direction (pink arrows,
Fig.~\ref{Fig2}). Similarly, electrons at $E_F$ in states dispersing along {\emph b}$^*$
can be thermally excited ($\Delta_C$) to the lowest-lying interchain conduction band above $E_F$ near
the X-point by absorbing phonons with opposite momenta. These phonons, with momenta $q\lesssim \sqrt{2}\pi/2b$
and energies $\hbar q v\lesssim 6$~meV (assuming a dispersionless, acoustic phonon with velocity $v=3$~km/s), will be excited in large
numbers at room temperature and above. The activation energies, from averaging the various band structure calculations,
are $\Delta_V=0.16$~eV and $\Delta_C=0.20$~eV. With $\Delta_V<\Delta_C$
a $p$-type thermopower should result.

These observations motivate an interpretation of the interchain transport in LiPB that reflects
parallel conduction through band-like states ($lo$), predominant at $T\lesssim 140$~K, and a thermally
activated, semiconductor contribution ($hi$), predominant at $T > 300$~K,
\begin{eqnarray}
\nonumber 
  \sigma&=&\sigma^{lo}+\sigma^{hi} \\
  S&=&\left(\sigma^{lo}/\sigma\right) S^{lo}+\left(\sigma^{hi}/\sigma\right) S^{hi}.
\end{eqnarray}
\begin{table}[b]
\caption{Fitting parameters for the semiconductor, high-$T$
component ($hi$) of the parallel conduction model discussed in the
text.}
\begin{ruledtabular}
\begin{tabular}{cccccccc}
Specimen  & $C (\Omega^{-1}\ {\rm cm}^{-1})$ & $E_{\sigma}$~(eV) & $D$ & $E_S$~(eV)\\
A& $2.40\times 10^3$ & 0.155 & -3.00 & 0.220\\
B& $2.41\times 10^3$ & 0.159 & -2.90 & 0.250\\
C& $1.60\times 10^3$ & 0.146 & -2.68 & 0.241\\
D& $1.85\times 10^3$ & 0.146 & -2.15 & 0.241\\
\end{tabular}
\end{ruledtabular}
\end{table}

\noindent The low-$T$ TEP is taken as $S^{lo}=32\ \mu$V/K,
independent of $T$ as motivated by the $S_c(T)$ data, and
$\sigma^{lo}=1/(A+BT)$ from linear-least-squares fits (dashed
lines, Fig.~\ref{Fig1}) to the $\rho_c(T)$ data in the range
40~K$\leq T\leq 140$~K. For the high-$T$ contribution we first
tried a single semiconductor component to minimize the number of
free parameters: $\sigma^{hi}=C\exp(-E_{\sigma}/k_BT)$ and
$S^{hi}=(k_B/|e|)(E_S/k_BT+D)$, where $C$ and $D$ are constants
and $E_{\sigma}$ and $E_S$ are activation energies. The solid
curves through the $\rho_c(T)$ and $S_c(T)$ data (Fig.~\ref{Fig1})
demonstrate the agreement achieved with Eq.~(1) throughout the
range $T\geq 40$~K using these simple forms for $\sigma^{hi}$ and
$S^{hi}$ (parameter values are listed in Table~I).  The
discrepancy between the computed and measured TEPs in the
transition region (near 200~K) may reflect our neglect of a
tunneling contribution, like that observed for the cuprates
[Fig.~\ref{Fig1}~(b)], but has negligible impact on the fitting at
high-$T$ if, as in the latter materials, this contribution adds a
small constant. The average activation energies found for the four
crystals (with s.d. uncertainties) are: $E_{\sigma}=0.15\pm
0.01$~eV and $E_S=0.24\pm 0.01$~eV.

The observation $E_S>E_{\sigma}$ is incompatible with single-band
($E_S=E_{\sigma}$) and intrinsic two-band ($E_S\leq E_{\sigma}$)
semiconductor conduction for the high-$T$ component, but is
naturally explained \cite{SM}, consistent with the band structure,
if both valence and conduction band states are excited with {\it
differing} activation energies, $\Delta_V$ and $\Delta_C$.
Analyzing the data with a three-component model, low-$T$ metallic
and two high-$T$ semiconducting contributions \cite{SM}, yields
average energies $\Delta_V\simeq 0.14$~eV and $\Delta_C\simeq
0.23$~eV, i.e. nearly the same as $E_{\sigma}$ and $E_S$ in the
two-component model. The
interband transitions evidently serve as the predominant mechanism
for interchain ({\emph c}-axis) transport above 400~K where the
semiconducting components represent more than 50\% of the total
conductivity \cite{SM}. In this regime LiPB thus behaves
simultaneously as a {\emph p}-type semiconductor and {\emph
n}-type metal along mutually perpendicular directions, leading to
its high Seebeck anisotropy.
\begin{figure}[b]
\includegraphics[width=3.25in,clip]{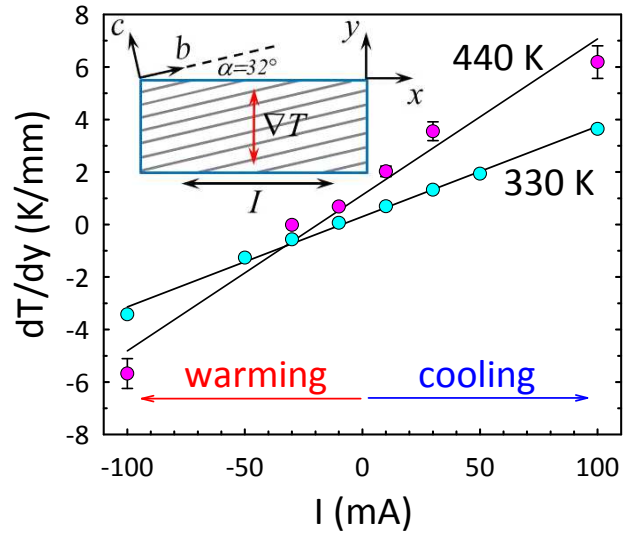}
\caption{(color online) Current dependence of the transverse
Peltier-induced temperature gradient for a LiPB crystal at
$T=330$~K and $440$~K. Error bars reflect uncertainty due to slow
oscillations in the Cu block temperature ($\leq 0.1$~K),
particularly at the highest $T$. The inset shows the orientation
of current and heat flow relative to the specimen (\emph{x-y}) and
crystallographic (\emph{b-c}) axes.}
\label{Fig3}
\end{figure}
\begin{figure}[t]
\includegraphics[width=3.25in,clip]{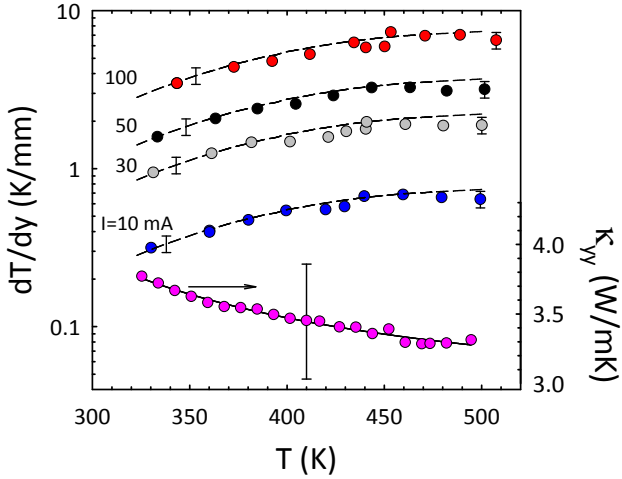}
\caption{(color online) $dT/dy$ (averaged for $+$,
$-$ current) {\emph vs.} the average specimen temperature at fixed
applied currents (left ordinate), and $\kappa_{yy}(T)$ (right
ordinate). The dashed curves are computed (see text) using
$\kappa_{yy}$ and $\Delta S$ interpolated from Fig.~\ref{Fig1}~(a). Error bars
for the data and computed curves (12\%) are determined by
uncertainty in the thermocouple junction separation. The
$\kappa_{yy}$ data were corrected for radiation losses and
conduction through the leads estimated from direct measurements on
similar specimens suspended from their leads. These corrections
amounted to 6\% at 300~K and 30\% at 500~K.}
\label{Fig4}
\end{figure}

To test for the transverse Peltier effect, a rectangular specimen
($x\times y\times z= 1.9\times 0.9 \times 0.2$~mm$^3$) with its
long axis at an angle $\alpha=32^{\circ}$ to the {\emph b} axis
(Fig.~\ref{Fig3}, inset), was mounted with one edge thermally
anchored to a copper heat sink in vacuum; current was applied
along the \emph{x} direction. Cooling or warming at the free edge
of the crystal, monitored by a differential thermocouple with
junctions along the $y$ direction (separation $d\simeq 0.4$~mm),
was induced by forward or reverse current, respectively, and was
linear in the current (Fig.~\ref{Fig3}). The averaged
temperature gradient, $dT/dy=[\Delta T(I+)-\Delta T(I-)]/2d$, is
shown as a function of the average
specimen temperature in Fig.~\ref{Fig4} for fixed values of
the current. The dashed curves in Fig.~\ref{Fig4} were
computed using $\Delta S$ [interpolated from data for crystal B,
Fig.~\ref{Fig1}~(a)], the transverse thermal conductivity
($\kappa_{yy}$) measured in a separate experiment with a heater
attached to the free end of the specimen (Fig.~\ref{Fig4},
right ordinate), and the heat flux equation \cite{Lengfellner},
$dT/dy=(TS_{xy}/\kappa_{yy})j$, where $S_{xy}=(1/2)\Delta
S\sin(2\alpha)$ and $j$ is the current density. The latter expression ignores a Joule heating
term (varying as $j^2$), justified by the linearity of $dT/dy$
with current noted above. Because LiPB has a rather low thermal
conductivity along the \emph{c} axis \cite{Nernst}, the transverse
thermoelectric figure of merit is among the largest known for a
single-phase material \cite{LargeZT},
$Z_{xy}T=TS_{xy}^2/(\rho_{xx}\kappa_{yy})\simeq 0.024\pm 0.007$ at
450~K \cite{ZTNote}.

LiPB may itself prove useful in converting waste heat to electrical power or in energy detection.
Given that the features underlying its extreme Seebeck anisotropy appear fairly generic -- low-dimensional, metallic
electronic structure and dispersing bands for the transverse direction in close proximity to $E_F$  -- the larger
implication from this study is that other materials with such properties may yet to be revealed.

The authors acknowledge M. Grayson and B. Cui for very helpful comments.
This material is based upon work supported by U.S. Department of Energy (DOE)/Basic Energy
Sciences (BES) Grant No. DE-FG02-12ER46888 (University of Miami), the National Science Foundation under grant
DMR-0907036 (Montana State University), and in Lorena by the CNPq (308162/2013-7) and FAPESP (2009/54001-2).

\newpage
\setcounter{figure}{0}

\renewcommand*{\thefigure}{S\arabic{figure}}
\renewcommand*{\theequation}{S\arabic{equation}}
\renewcommand*{\thetable}{S\arabic{table}}
\onecolumngrid
\newpage

\begin{center}
{\large\bf Supplementary Material: Extreme thermopower anisotropy and interchain transport in the quasi-one-dimensional metal Li$_{0.9}$Mo$_6$O$_{17}$}
\vskip 0.17in
J.~L.~Cohn, S.~Moshfeghyeganeh, C.~A.~M. dos Santos, J.~J.~Neumeier
\vskip 0.05in
\end{center}

\subsection*{Additional Data for $S_c$ and $\rho_c$}
\begin{figure}[h]
\includegraphics[width=4.in,clip]{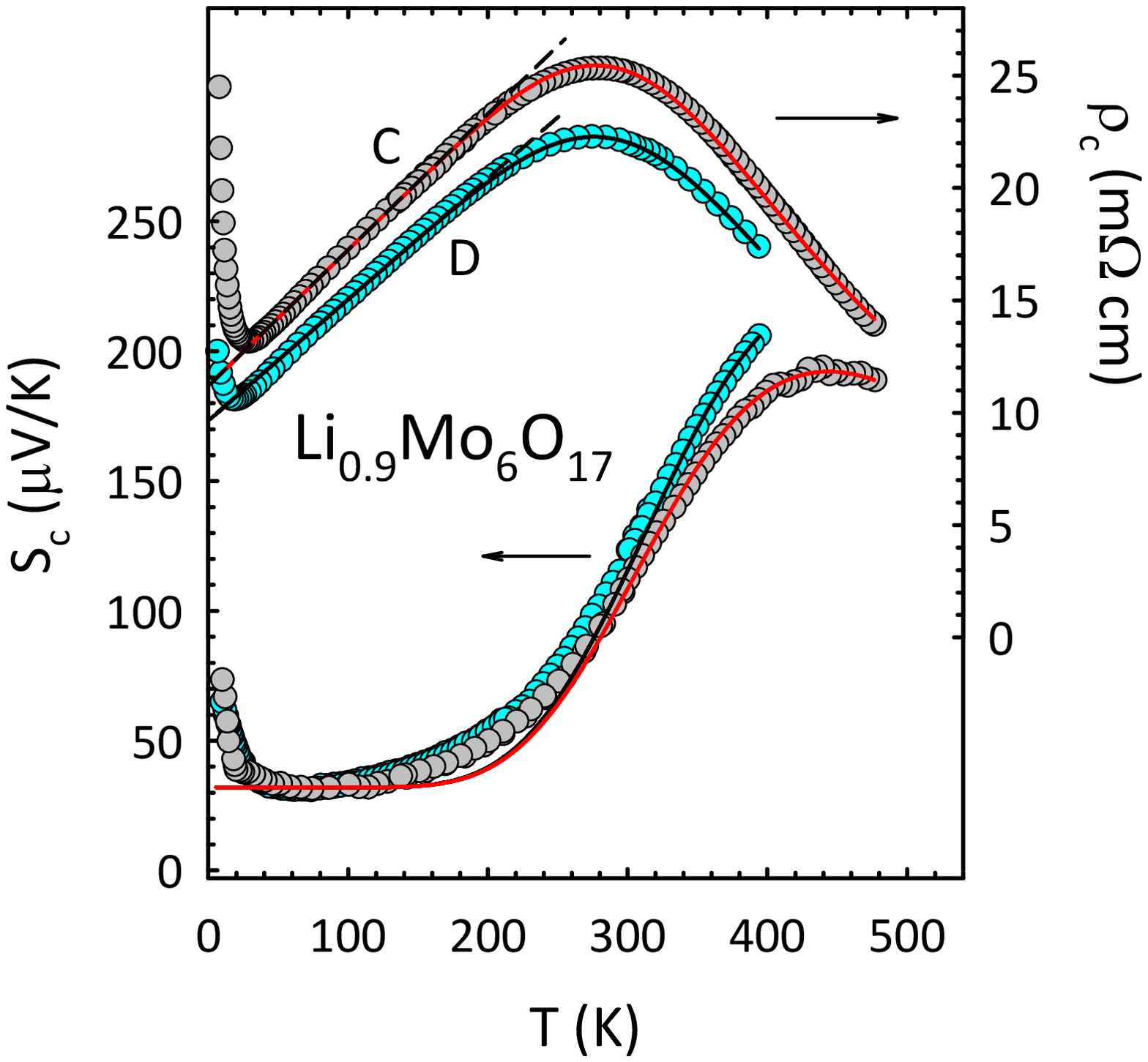}
\caption{LiPB inter-chain ({\emph c}-axis) thermopowers (left ordinate) and resistivities (right
ordinate) for crystals C and D. Solid curves through the data are fits to the parallel
conduction model discussed in the text; parameters for the semiconducting component are listed in Table~I.
Dashed lines are linear-least-squares fits to the low-$T$ $\rho_c(T)$.}
\label{S1}
\end{figure}

\subsection*{Differing excitation energies for valence and conduction bands yields $E_S>E_{\sigma}$}

Here we demonstrate that a single semiconductor component with $E_S>E_{\sigma}$,
as emerged from the two-component model (metal $+$ semiconductor)
discussed in the text, can be reproduced by a two-component semiconductor
with differing hole and electron excitation energies,
$\Delta_V=E_F-E_V$ and $\Delta_C=E_F-E_C$, respectively, and $\Delta_V<\Delta_C$.

\begin{figure}[b]
\includegraphics[width=4.in,clip]{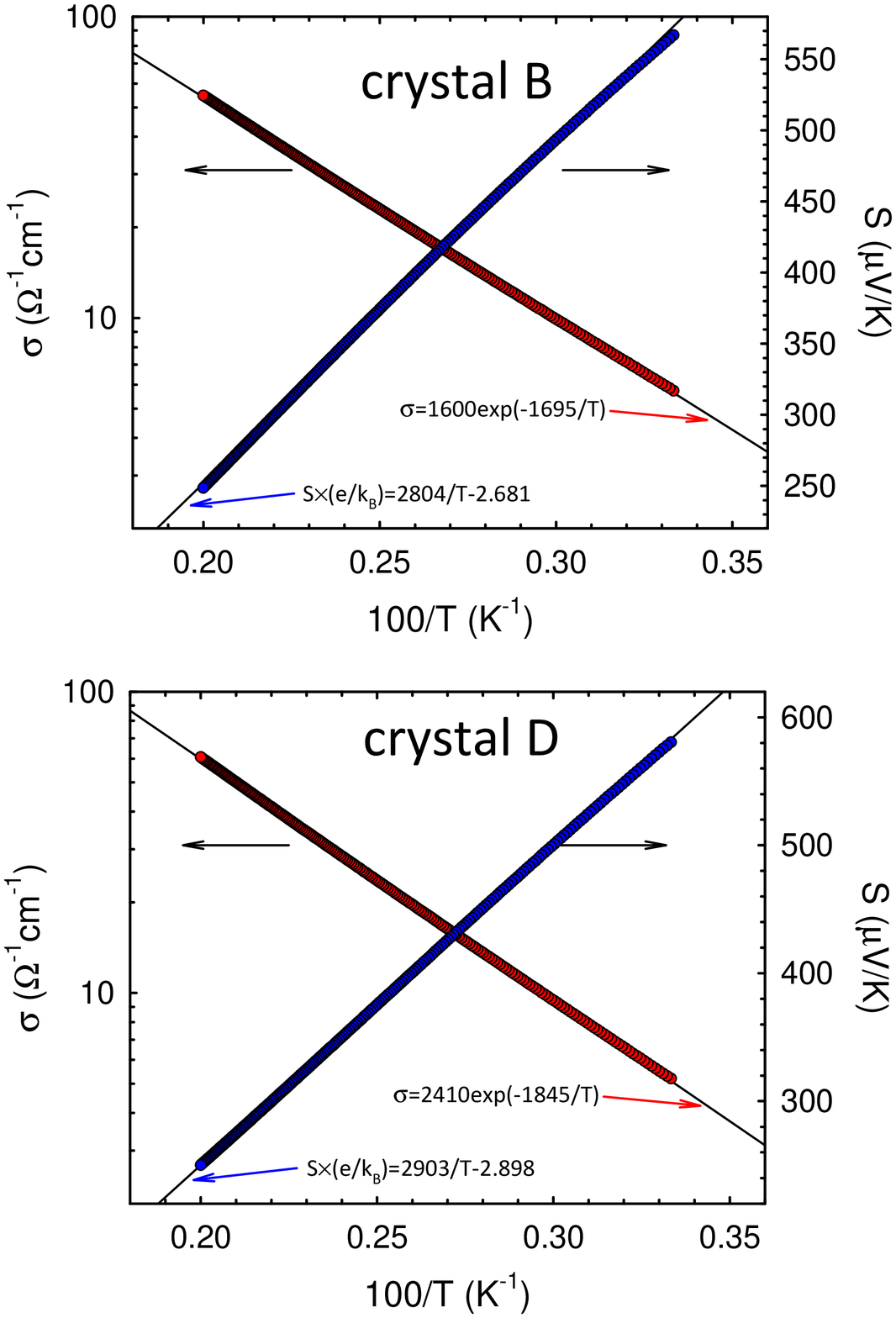}
\caption{Conductivity and thermopower computed from the
two-component semiconductor model having effective activation
energies (solid lines) matching those of the single-component
semiconductor contribution to the fitting to specimen B discussed
in the text (Table~I).} \label{S2}
\end{figure}

Distinguishing electron ($e$) and hole ($h$) contributions, we
have:
\begin{eqnarray}
\nonumber 
  \sigma&=&\sigma_e+\sigma_h= C_C\exp(\Delta_C/k_BT)+C_V\exp(\Delta_V/k_BT)\\
\nonumber
\\
\nonumber
 S_e&=&-{k_B\over e}\left({\Delta_C\over k_BT}+A_e\right);\ S_h={k_B\over e}\left({\Delta_V\over k_BT}+A_h\right)\\
\nonumber
\\
\nonumber
  S&=&\left(\sigma_e/\sigma\right) S_e+\left(\sigma_h/\sigma\right) S_h.
\end{eqnarray}
\noindent Figure~\ref{S2} shows $\sigma(T)$ and $S(T)$ computed
for the following parameters appropriate to crystal B:
$\Delta_C=0.219$~eV, $\Delta_V=0.135$~eV, $C_C=1900\
\Omega^{-1}{\rm cm}^{-1}$, $C_V=985\ \Omega^{-1}{\rm cm}^{-1}$,
$A_e=2.80$, and $A_h=2.69$. The solid lines demonstrate that this
set of parameters produces effective single-component parameters
identical to those of the single semiconductor component listed in
Table~I for specimen B and plotted in Fig.~1.
The constants $A_e$ and $A_h$ represent weighted averages over the
charge carriers in the conduction and valence
bands [38]. For example, $A=3$ corresponds to a density
of states and mobility that increase linearly with energy, and
$A=1$ for constant density of states and mobility. Temperature
dependent band energies, varying linearly in $T$ to lowest order,
can contribute to these constant terms and even alter their
sign [39].
\begin{table}[t]
\caption{Fitting parameters for the semiconducting components of the three-component model for three specimens.}
\begin{ruledtabular}
\begin{tabular}{cccccccc}
Specimen  & $C_C (\Omega^{-1}\ {\rm cm}^{-1})$ & $\Delta_C$~(eV) & $C_V (\Omega^{-1}\ {\rm cm}^{-1})$ & $\Delta_V$~(eV) & $A_e$ & $A_h$\\
B& $2.30\times 10^3$ & 0.231 & $1.40\times 10^3$  & 0.146 & 2.70 & 2.20\\
C& $2.00\times 10^3$ & 0.237 & $1.10\times 10^3$  & 0.137 & 2.62 & 1.74\\
D& $2.00\times 10^3$ & 0.224 & $1.32\times 10^3$  & 0.138 & 2.60 & 2.70\\
\end{tabular}
\end{ruledtabular}
\end{table}

\subsection*{Three-component fitting}

\begin{figure}[h]
\includegraphics[width=3.8in,clip]{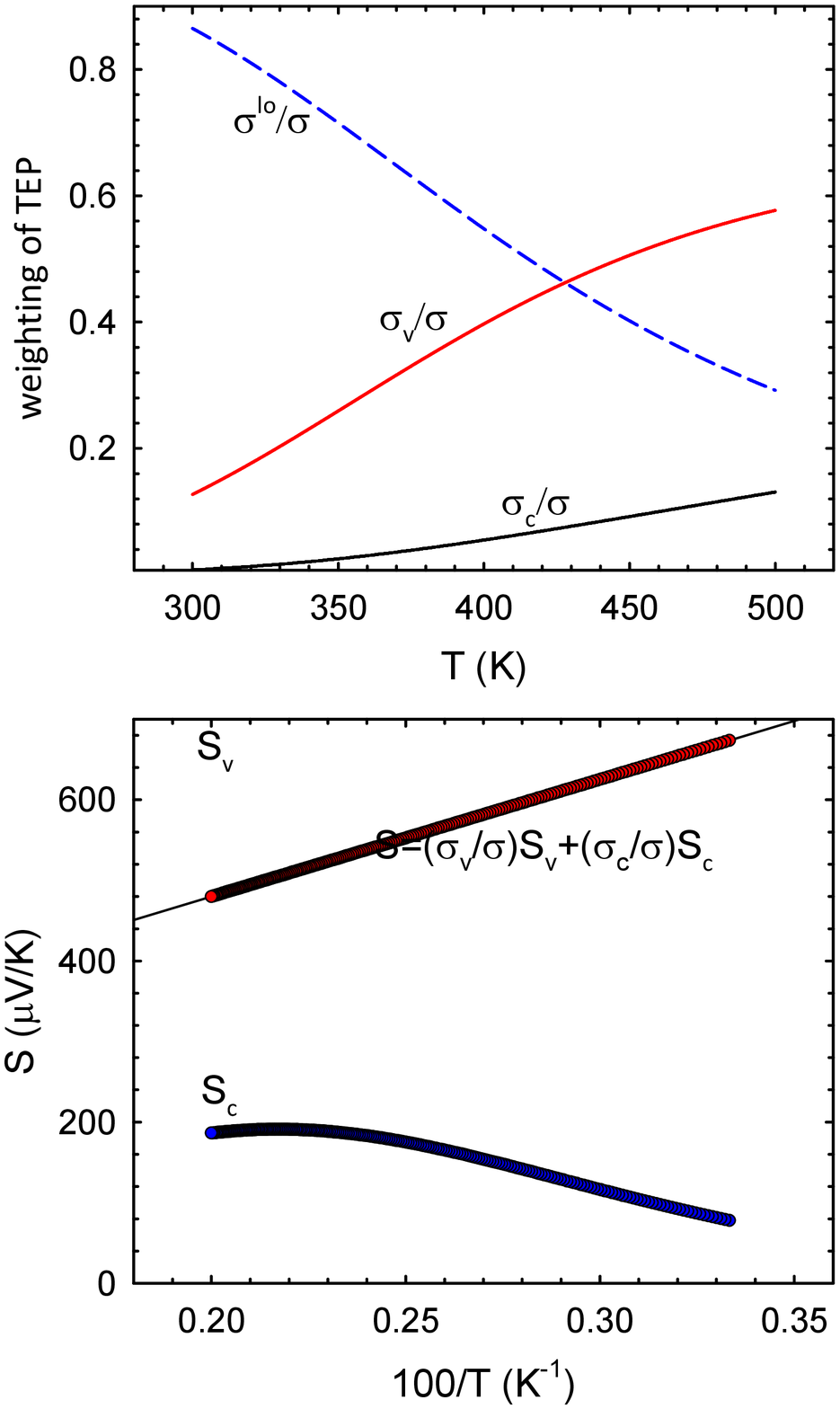}
\caption{$T$-dependent weights (conductivity ratios) for each of the three components from three-component fitting
to crystal B (parameters from Table~S1).}
\label{S3}
\end{figure}

Incorporating two semiconductor components along with the low-$T$ term described in the text
constitutes a three-component model with appropriate weighting by the respective partial conductivities.
This procedure produces fits that are indistinguishable from those of Fig.~1; Table~S1
lists fitting parameters describing the semiconducting components for three crystals. Crystal A was excluded
because its data do not extend to high enough temperature to sufficiently constrain the semiconducting parameters.
Note that this analysis yields values for $\Delta_C$ and $\Delta_V$ that are only 5-10\% smaller than $E_S$ and $E_{\sigma}$,
respectively, of the simpler (two-component) model (Table~I). Figure~\ref{S3} shows the $T$-dependent weights (fractional conductivities)
for each of the three components using the fitting parameters for crystal B.

\vfill\eject

\noindent
[38] H. Fritzsche, Sol. St. Commun. {\bf 9,} 1813 (1971).

\noindent
[39] V. A. Johnson and K. Lark-Horovitz, Phys. Rev. {\bf 92,} 226 (1953).


\begin{thebibliography}{39}
\bibitem{ZTNote} Since $\rho_{xx}$ was not measured
for the Peltier crystal, we computed it using the comprehensive results for $\rho_b$ from Wakeham {\it et al.}
[Nature Commun. {\bf 2,} 396 (2011)], linearly extrapolated to 450~K, and the $\rho_c$
data for crystal B [Fig.~1~(a)]. Using the appropriate tensor expressions (Ref.~2), this gave $\rho_{xx}\simeq 4.62\times 10^{-5}\ \Omega\ {\rm m}$.
The linearity in $\Delta T (I)$ (Fig.~3) implies that the actual $\rho_{xx}$ for our specimen cannot be
much larger than this value (we estimate not by more than 30\%); a deviation from linearity (suppressed $\Delta T$) that is quadratic
in $j$ would be the signature of Joule heating from the second term in the heat flux equation [2].
\end{thebibliography}
\end{document}